\documentclass[aps,twocolumn,superscriptaddress,showpacs,floatfix,nofootinbib,superscriptaddress,]{revtex4-1}
\usepackage{epsfig,amsmath,amssymb,bm,color,graphicx}
\usepackage{pdfpages}
\usepackage{afterpage}

\def\bea{\begin{eqnarray}}
\def\eea{\end{eqnarray}}
\def\be{\begin{equation}}
\def\ee{\end{equation}}

\newcommand{\de}{\mathrm d}

\newcommand{\odm}{{\Omega_\mathrm{DM}}}

\newcommand{\mdm}{{m_{\rm DM}}}

\newcommand{\sv}{{\langle\sigma_a v\rangle}}

\newcommand{\g}{$\gamma$}

\newcommand{\low}{{\sc low}}
\newcommand{\high}{{\sc high}}

\begin{document}
\setcounter{tocdepth}{3}

\title{Particle dark matter searches outside the Local Group}

\author{Marco Regis}
\email{regis@to.infn.it}
\affiliation{Dipartimento di Fisica, Universit\`{a} di Torino and Istituto Nazionale di Fisica Nucleare, Sezione di Torino, via P. Giuria 1, I--10125 Torino, Italy}
\author{Jun-Qing Xia}
\email{xiajq@ihep.ac.cn}
\affiliation{Key Laboratory of Particle Astrophysics, Institute of High Energy Physics, Chinese Academy of Science, P. O. Box 918-3, Beijing 100049, P. R. China}
\affiliation{Collaborative Innovation Center of Modern Astronomy and Space Exploration, P. R. China}
\author{Alessandro Cuoco}
\email{cuoco@to.infn.it}
\affiliation{Dipartimento di Fisica, Universit\`{a} di Torino and Istituto Nazionale di Fisica Nucleare, Sezione di Torino, via P. Giuria 1, I--10125 Torino, Italy}
\author{Enzo Branchini}
\affiliation{ Dipartimento di Matematica e  Fisica, Universit\`a degli Studi ``Roma
Tre'', via della Vasca Navale 84, I-00146 Roma, Italy}
\affiliation{INFN, Sezione di Roma Tre, via della Vasca Navale 84, I-00146 Roma, Italy}
\affiliation{INAF Osservatorio Astronomico di Roma, INAF, Osservatorio Astronomico di Roma, Monte Porzio Catone, Italy}
\author{Nicolao Fornengo}
\affiliation{Dipartimento di Fisica, Universit\`{a} di Torino and Istituto Nazionale di Fisica Nucleare, Sezione di Torino, via P. Giuria 1, I--10125 Torino, Italy}
\author{Matteo Viel}
\affiliation{INAF Osservatorio Astronomico di Trieste, Via G. B. Tiepolo 11,
I-34141, Trieste, Italy}
\affiliation{INFN, Sezione di Trieste, via Valerio 2, I-34127, Trieste, Italy}


\begin{abstract}
If dark matter (DM) is composed by particles which are non-gravitationally coupled to ordinary matter, their annihilations or decays in cosmic structures can result in detectable radiation. We show that the most powerful technique to detect a particle DM signal outside the Local Group is to study the angular cross-correlation of non-gravitational signals with low-redshift gravitational probes. This method allows to enhance signal-to-noise from the regions of the Universe where the DM-induced emission is preferentially generated. We demonstrate the power of this approach by focusing on GeV-TeV DM and on the recent cross-correlation analysis between the 2MASS galaxy catalogue and the Fermi-LAT gamma-ray maps. We show that this technique is more sensitive than other extragalactic gamma-ray probes, such as the energy spectrum and angular autocorrelation of the extragalactic background, and emission from clusters of galaxies. Intriguingly, we find that the measured cross-correlation can be well fitted by a DM component, with thermal annihilation cross section and mass between 10 and 100 GeV, depending on the small-scale DM properties and gamma-ray production mechanism. This solicits further data collection and dedicated analyses.
\end{abstract}

\maketitle

\section{Introduction}
\label{sec:intro}
The origin of cosmic structures is well understood in terms of evolution of matter perturbations arising after the inflationary period. Inhomogeneities starting off with higher-than-average density  grow through gravitational instability. DM is a necessary ingredient to the process, as it provides the potential-wells where standard matter is accreted after decoupling and protohalos form. As structure formation evolves, DM halos of increasing size form in a bottom-up fashion.

If DM is in form of particles which exhibit non-gravitational couplings to ordinary matter, a certain level of emitted radiation is expected. Photons can be produced from interactions of DM with the ambient medium (e.g., through scatterings) or from DM annihilation or decay by means of direct emission or through the production of intermediate particles. The non-gravitational signal associated to decay is proportional to the DM density: it is stronger at low redshift, because the produced radiation is diluted by the expansion of the Universe more rapidly than its source, i.e. the DM particle density. The DM annihilation signal, which is proportional to the density squared, is also peaked at low redshift since the density contrast associated to cosmic structures grows nonlinearly.

DM constitutes the backbone of all cosmic structures and DM halos represent, collectively, a potential source of DM decay or annihilation signals. This means that even if the radiation originating from DM annihilations or decays in a single halo is too faint to be detected, their {\sl cumulative signal} and its {\sl spatial coherence} could be.
In addition, since the DM signal is expected to peak at  $z<0.3$, it can be separated by more mundane astrophysical processes that typically trace the star formation history and peak at higher redshifts.

To increase the sensitivity to non-gravitational DM sources one needs to isolate the annihilation/decay signal  produced at low redshift. An effective way to  filter out any signal that is not associated to DM-dominated structures or that is originated at high redshift is to {\sl cross-correlate} the radiation field with {\it bona fide} low-redshift DM tracers \cite{Camera:2012cj,Ando:2013xwa,Fornengo:2013rga,Shirasaki:2014noa,Ando:2014aoa,Camera:2014rja}.
In the following, we adopt this approach in the specific and yet very relevant framework of weakly interacting massive particles (WIMP) that may either annihilate or decay. 
We will use the results of the cross-correlation analysis between \g-ray maps from Fermi-LAT \cite{Atwood:2009ez} and the 2MASS catalogue of relatively nearby galaxies~\cite{2MASS} presented in \cite{Xia:2015wka}.


\section{Data and Models}
\label{sec:data}
The cross angular power spectrum (CAPS) between the {\sl unresolved} \g-ray sky observed by Fermi-LAT and the distribution of 2MASS galaxies can be written as \cite{Fornengo:2013rga}:
\be
 C_\ell^{(\gamma g)}=\int \frac{d\chi}{\chi^2} W_{\gamma}(\chi)\, W_{g}(\chi)\,P_{\gamma g}\left(k=\ell/\chi,\chi\right)\;,
\label{eq:clfin}
\ee
where $\chi(z)$ denotes the radial comoving distance, $W_i(\chi)$ represent the window functions described below, $P_{\gamma g}(k,z)$ is the three-dimensional cross power spectrum (PS),  $k$ is the modulus of the wavenumber, and $\ell$ is the multipole. Indices \g\ and $g$ refer to $\gamma$-ray emitters and extragalactic sources in 2MASS, respectively. In Eq.~(\ref{eq:clfin}) we used the Limber approximation~\cite{limber}, since $P_{\gamma g}$ varies (relatively) slowly with $k$.

The (differential in energy) window function for \g-ray emission from  DM annihilation $W_{\gamma}(z)$ is \cite{Fornengo:2013rga}:
\be
W_{\gamma}^{a}(z) = \frac{(\odm \rho_c)^2\,\sv}{8\pi\,\mdm^2} 
 \left(1+z\right)^3 
\Delta^2(z) \, \frac{\de N_a}{\de E_\gamma}\,
 e^{-\tau\left[z,E_\gamma(z)\right]},
\label{eq:window_annDM}
\ee
where $\odm$ is the DM\footnote{A 6-parameter flat $\Lambda$CDM cosmological model is assumed with the value of the parameters taken from Ref.~\cite{Planck:2015xua}.} mean density in units of the critical density  $\rho_c$,  $\Delta^2(z)$ is the clumping factor, $\mdm$ is the mass of the DM particles, and $\sv$ denotes the velocity-averaged annihilation rate.
$\de N_a / \de E_\gamma$ indicates the number of photons produced per annihilation and determines the \g-ray energy spectrum. The exponential damping quantifies the absorption due to extra-galactic background light~\cite{Franceschini:2008tp}. 

The window function for DM decay is \cite{Fornengo:2013rga}:
\be
W_{\gamma}^{d}(z) = \frac{\odm \rho_c\,\Gamma_{\rm d}}{4\pi\,\mdm} 
\, 
\frac{\de N_d}{\de E_\gamma} \,
e^{-\tau\left[z,E_\gamma(z)\right]}\;,
\label{eq:window_decDM}
\ee
where $\Gamma_{\rm d}=1/\tau_d$ is the DM decay rate.

The window function of 2MASS galaxies is $W_{g}(z)\equiv H(z)/c\,dN_g/dz$ and their redshift distribution $dN_g/dz$ is \cite{Xia:2011}:
\be 
\frac{dN_g}{dz}(z)=\frac{\beta}{\Gamma(\frac{m+1}{\beta})}\frac{z^m}{z^{m+1}_0}
\exp\left[-\left(\frac{z}{z_0}\right)^\beta\right]\,,
\label{eq:dndz2mass}
\ee
with $m=1.90$, $\beta=1.75$ and $z_0=0.07$.

We employed the 2MASS catalogue instead of other compilations because the galaxy distribution in Eq.~\ref{eq:dndz2mass} is peaked at very low redshift as for the DM emission of Eqs.~\ref{eq:window_annDM} and \ref{eq:window_decDM}. This enhances the cross-correlation signal. The picture for astrophysical components would be different and other catalogues might be more informative (see Section S2 in the Supplemental Material~\cite{Supplemental}).

The PS $P_{\gamma g}$ in Eq.~(\ref{eq:clfin}) is computed within the halo-model framework, as the sum of a one-halo plus a two-halo terms. For more details, see~\cite{Fornengo:2013rga}. Both the PS and the clumping factor $\Delta^2(z)$ in Eq.~(\ref{eq:window_annDM}) depend on a number of DM properties: the halo mass function, that we take from Ref.~\cite{Sheth:1999mn}, the halo density profile, for which we assume a Navarro-Frenk-White model \cite{Navarro:1996gj}, the minimum halo mass, that we set equal to $10^{-6} M_\odot$, and the halo mass-concentration relation $c(M,z)$, that we adopt from Ref.~\cite{Prada:2011jf}. The theoretical uncertainty of these quantities is rather small for halos larger than $10^{10}\,M_\odot$, because they can be constrained by observations and simulations. Since the DM decay signal is mainly contributed by large structures, the theoretical predictions are relatively robust. This is not the case for the annihilation signal which is preferentially produced in small halos and in substructures within large halos. Consequently,  theoretical uncertainties on the annihilation signal are larger. For the subhalo contribution we consider two scenarios (\low\ and \high) to bracket theoretical uncertainty. The \low\ case follows the model of Ref.~\cite{Sanchez-Conde:2013yxa} (see their Eq.~(2), with a subhalo mass function $dn/dM_{\rm sub}\propto M_{\rm sub}^{-2}$). The \high\ scenario is taken from Ref.~\cite{Gao:2011rf}, with  the halo mass-concentration relation extrapolated down to low masses as a power law.
Further uncertainties in the concentration parameter and in the value of the minimum halo mass can introduce an extra factor of $\sim2$ of uncertainty, that we quantify only in the Supplemental Material~\cite{Supplemental}, for the sake of definiteness.

\begin{figure}[t]
\vspace{-4cm}
\centering
\includegraphics[width=0.55\textwidth]{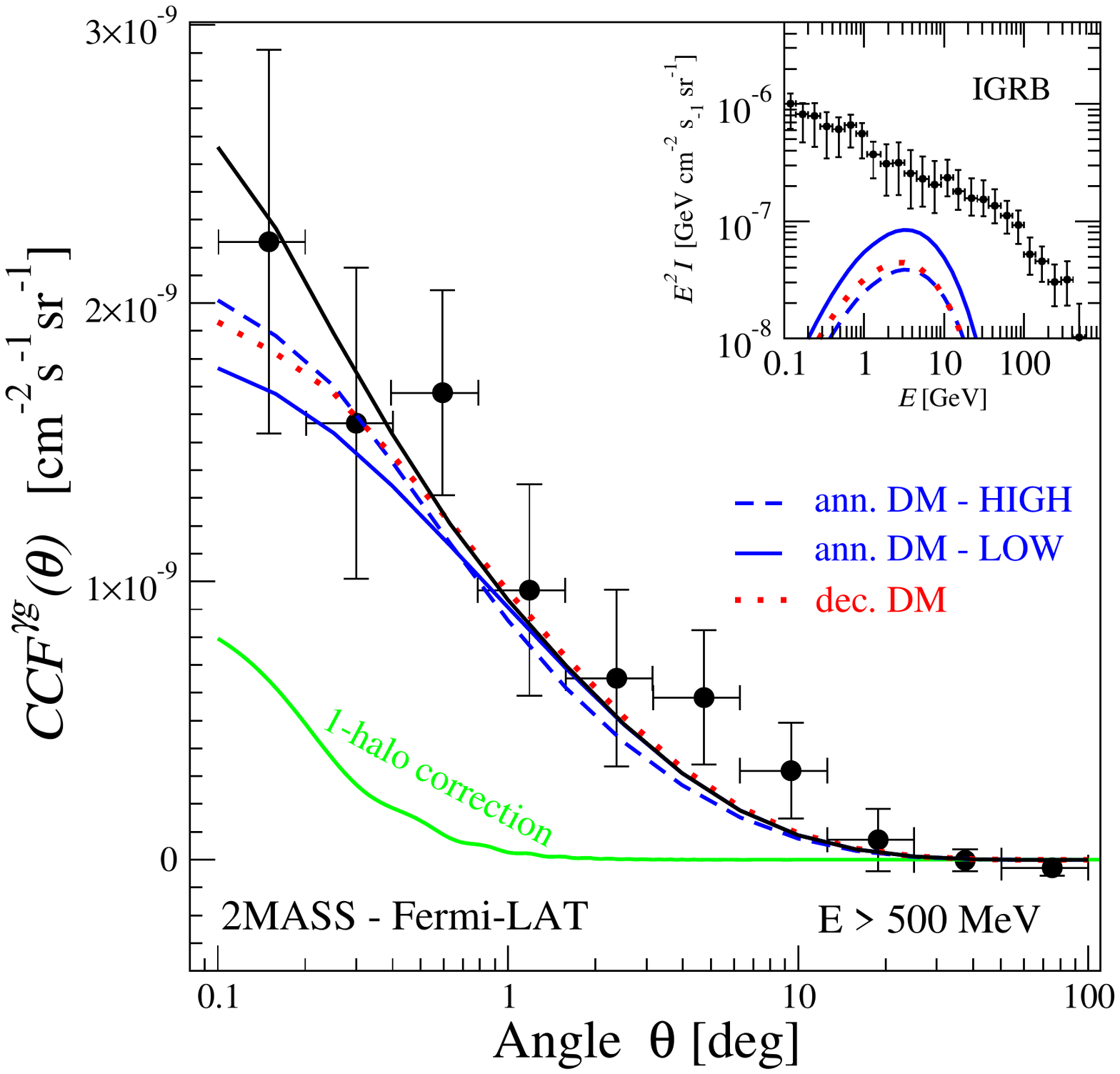}
\caption{Cross-correlation above 500 MeV for the best fitting annihilating and decaying DM scenarios, compared to the measured CCF. The curves are for DM particles of 100 GeV (200 GeV) annihilating (decaying) into $b\bar b$. We show the two annihilation models \high\ and \low\ with annihilation rates $\sv=2 \times 10^{-26}\,\mathrm{cm^3s^{-1}}$ (blue-dashed)  and $2.4 \times 10^{-25}\,\mathrm{cm^3s^{-1}}$ (blue-solid), respectively, and a decay model with lifetime $\tau=1.6 \times 10^{27}$ s (red-dotted). The green curve shows the CCF of the 1-halo correction term $C_{1h}$. We show the sum of this component and the DM CCF (in the \low\ scenario) with the black curve. The inset shows that these DM models provide a subdominant contribution to the observed IGRB spectrum~\cite{Ackermann:2014usa}.}
\label{fig:DMfit}
\end{figure}

\begin{figure*}[t]
\centering
\includegraphics[width=0.49\textwidth]{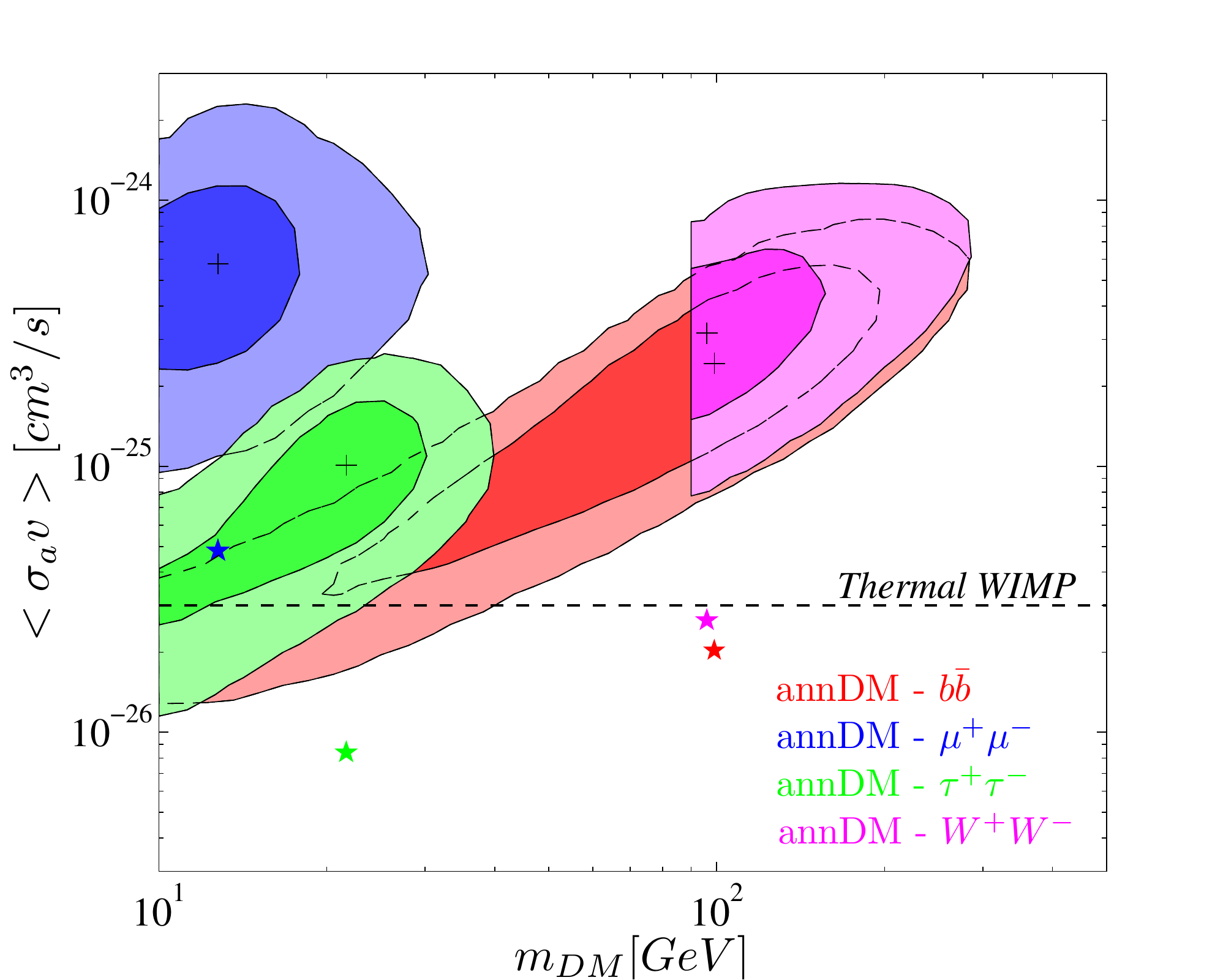}
\includegraphics[width=0.49\textwidth]{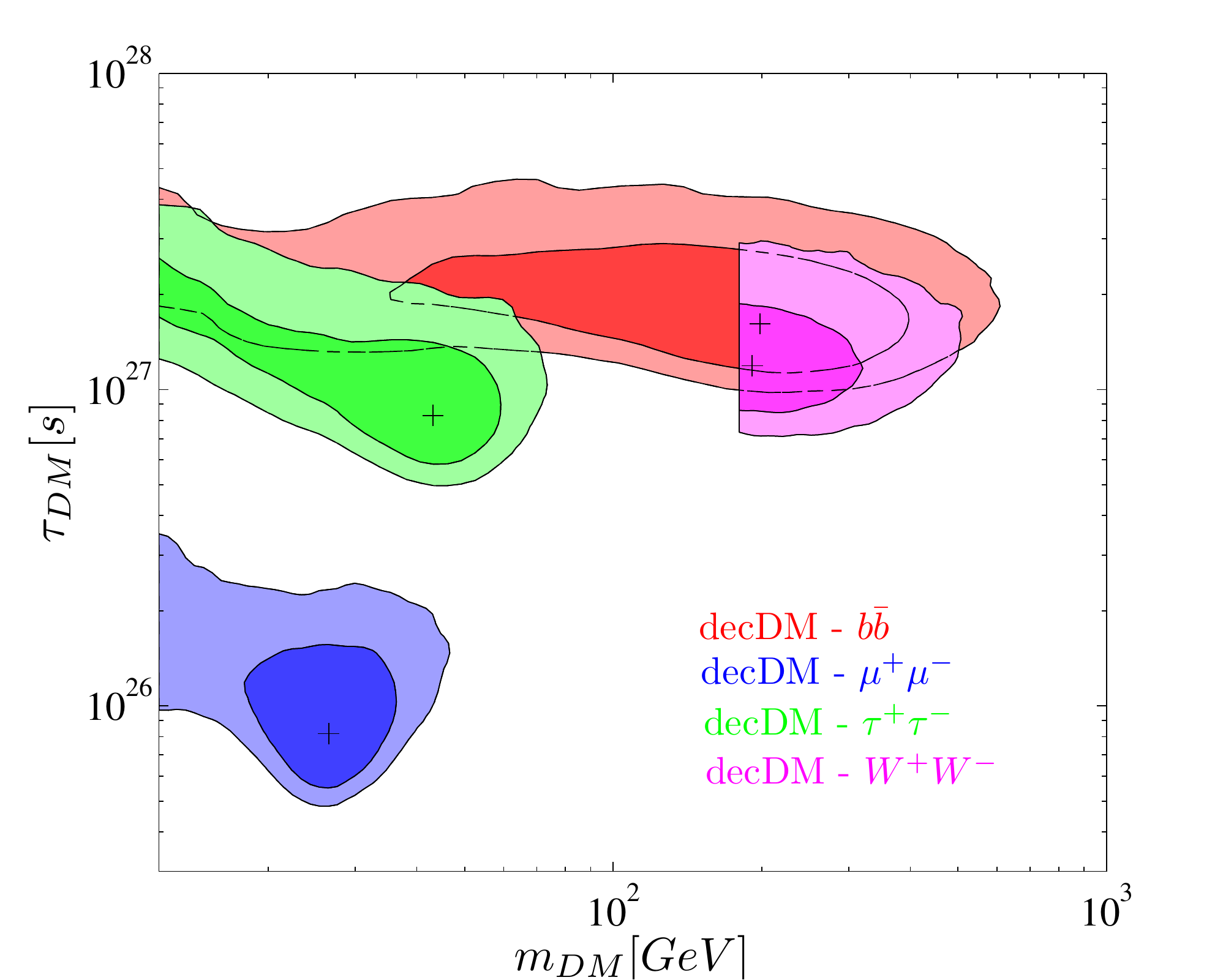}
\caption{Left: 1$\sigma$ and 2$\sigma$ allowed regions for the DM annihilation rate versus its mass, for different \g-ray production channels and assuming a \low\ substructure scheme. Crosses indicate the best fitting models. In the \high\ scenario, regions remain similar but shifted downward by a factor of $\sim12$, see stars indicating the best fitting models. Right: The same as in the left-hand panel but for decaying DM, showing the DM particle lifetime as a function of its mass. }
\label{fig:DMtree}
\end{figure*}

In our CAPS model (Eq.~\ref{eq:clfin}), we add a constant term $C_{1h}$ ({\it one-halo correction term}) to correct for possible unaccounted correlations at very small-scales, within the Fermi-LAT Point Spread Function (PSF).
The value of $C_{1h}$ will be determined by fitting the data, and 
we anticipate that we find a $C_{1h}$ value compatible with zero. Thus, the inclusion of this term does not change significantly the results.
For a discussion on this term, see Ref.~\cite{Ando:2014aoa}.

The measured CAPS $\tilde C_\ell^{(\gamma g)}$ is a convolution of the true CAPS and the effective beam window function $W_\ell^{B}$ that accounts for the PSF of the instrument and the pixelization of the $\gamma$-ray map. Both quantities depend on energy. 
We use the  $W_\ell^{B}$ derived in Ref.~\cite{Xia:2015wka} and model the observed spectrum as $\tilde C_\ell^{(\gamma g)}=W_\ell^{B}\,C_\ell^{(\gamma g)}$.

In the following, we shall consider the angular cross correlation function (CCF) rather than the spectrum. To model the CCF, we Legendre-transform the CAPS:
\be
 CCF^{(\gamma g)}(\theta) = \sum_\ell\frac{2\ell+1}{4\pi}\tilde C_\ell^{\gamma g} P_\ell[\cos(\theta)] \;,
\label{eq:2point}
\ee
where $\theta$ is the angular separation and $P_\ell$ are the Legendre polynomials.

To compare model and observed CCFs, we estimate the $\chi^2$ difference defined as:
\be
   \chi^2 = \sum_{n=1}^3\,\sum_{\theta_i\,\theta_j}   \left(d_{\theta_i}^{n} -m_{\theta_i}^{n}(\bm A) \right) \, \left[C^{n}\right]^{-1}_{\theta_i\theta_j}  \left(d_{\theta_j}^{n} -m_{\theta_j}^{n}(\bm A) \right) \, ,
\label{eq:chi2}
\ee
where $m$ and $d$ indicate model and data, $n$ identifies each one of the three overlapping energy ranges considered ($E>0.5 \,, 1, \, {\rm and} \, 10$ GeV) and the indices $\theta_i$ and $\theta_j$ run over 10 angular bins logarithmically spaced between $\theta= 0.1^\circ$ and $100^\circ$. $C_{\theta_i\theta_j}^{n}$ is the covariance matrix that quantifies the errors of the data and their covariance among the angular bins.  Data and covariance matrix are taken from Ref.~\cite{Xia:2015wka}. The parameter vector for annihilating DM is
$\bm A = \left[\mdm,\,\sv,\,C_{1h}\right]$, whereas for the decaying DM is $\bm A=\left[\mdm,\,\tau_d,\,C_{1h}\right]$.

\section{results}
\label{sec:results}

\begin{figure*}[t]
\centering
\includegraphics[width=0.49\textwidth]{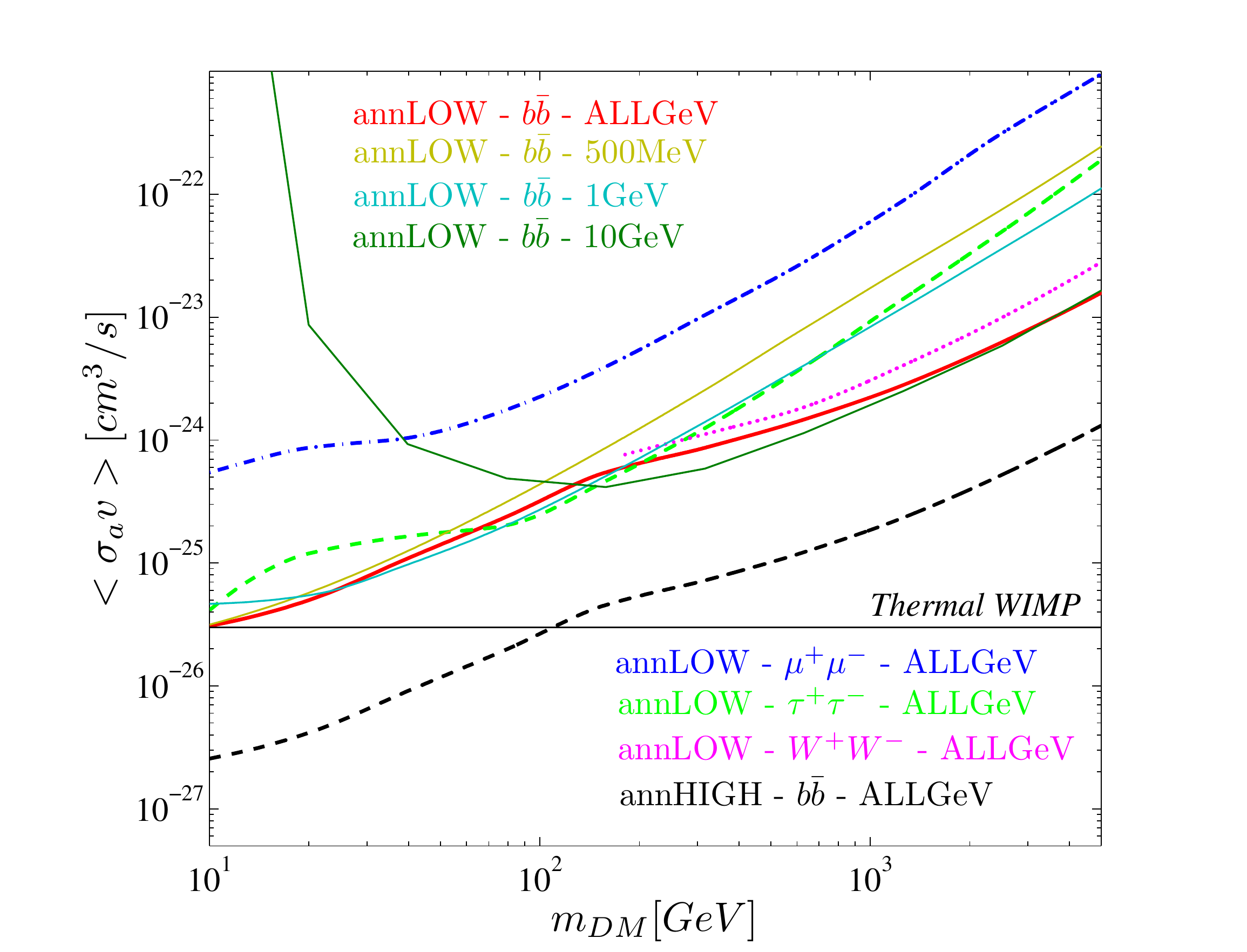}
\includegraphics[width=0.49\textwidth]{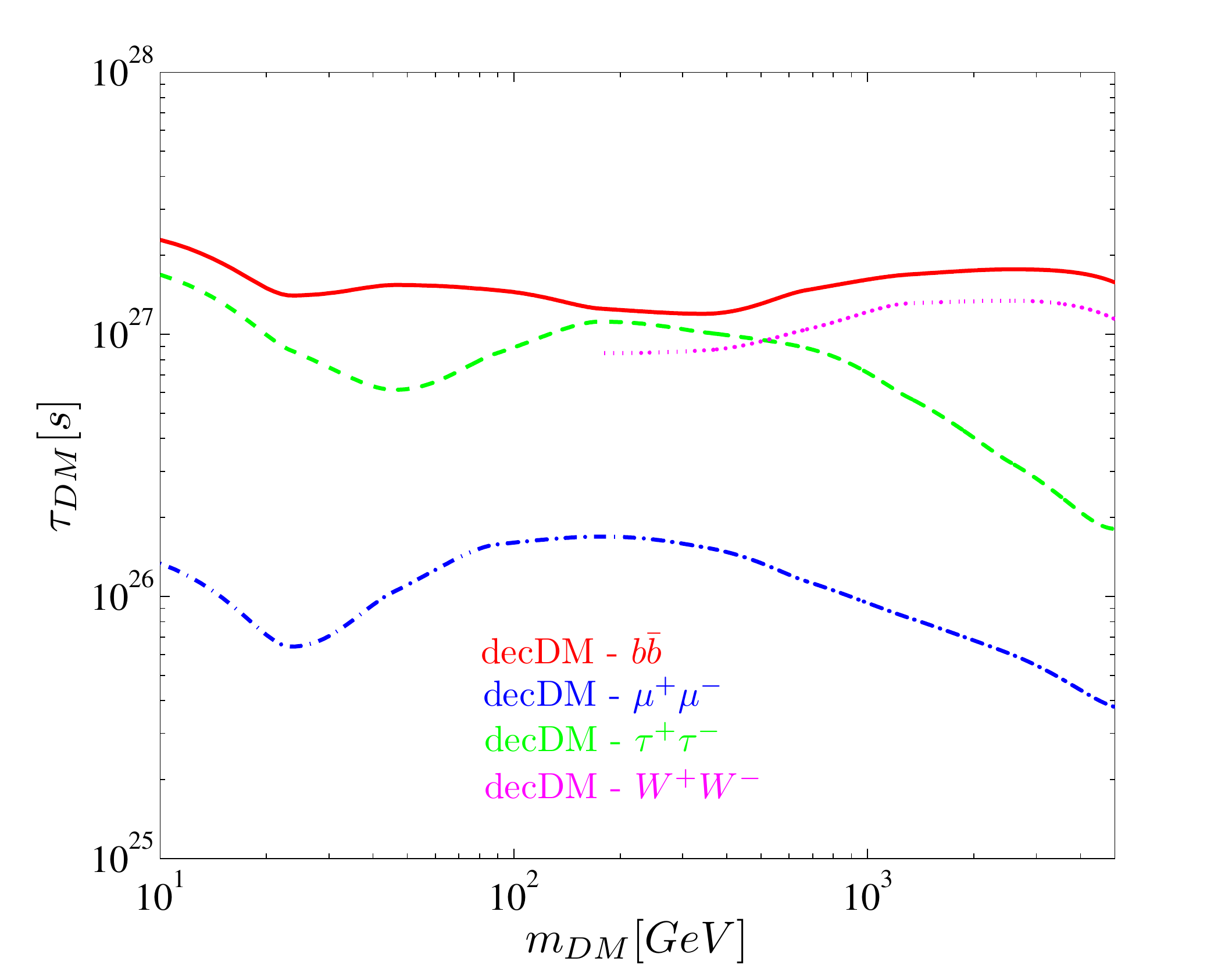}
\caption{Left: 95\% C.L. upper limits on the DM annihilation rate as a function of its mass. Both \high\ and \low\ clustering schemes are shown for WIMPs annihilating into $b\bar b$ (with the impact of different energy bins reported for the latter case). Other final states of annihilation ($\mu^+\mu^-$, $\tau^+\tau^-$, $W^+W^-$) are shown in the \low\ scenario only, for clarity. Right: 95\% C.L. lower limits on the DM lifetime as a function of its mass, for different final states of decay.}
\label{fig:DMbounds}
\end{figure*}

In Fig.~\ref{fig:DMfit} we show a comparison between the measured CCF in one of the considered energy bins ($E>500$ MeV) and the best fitting annihilating and decaying DM models obtained from the analysis discussed below. Errorbars are given by the diagonal elements of the covariance matrix. DM models fit the measured CCF remarkably well (for best fitting model, $\chi^2_{BF}=16.7$ with 26 d.o.f.). It is also noteworthy that the level of annihilation/decay rate provides a minor contribution to the Isotropic Gamma Ray Background (IGRB) measured by the Fermi-LAT~\cite{Ackermann:2014usa}, as shown in the inset of the figure. This implies that the cross-correlation technique can detect a DM signals too faint to show up in the total intensity measurement (for a review of the IGRB properties, see Ref.~\cite{Fornasa:2015qua}).

In Fig.~\ref{fig:DMtree}, we show the 1$\sigma$ and 2$\sigma$ C.L. contours (obtained marginalizing over $C_{1h}$) for DM mass and annihilation/decay rate for various final states.
Note that, although we use only three energy bins, they are sufficient to constrain the DM mass which induces a small but characteristic signature  in the energy spectrum. In the \low\ scenario the $1\sigma$ region lies just above the thermal annihilation rate $\sv=3 \times 10^{-26}\,\mathrm{cm^3s^{-1}}$. In the \high\ case, the DM signal increases by a factor of $\sim$ 10 and consequently regions shift down by one order of magnitude. Therefore, given the current uncertainty in modeling DM structures we conclude that the thermal cross section is well within the allowed regions for $m_{DM}\lesssim 200$ GeV.

We stress that the confidence contours in Fig.~\ref{fig:DMtree} are drawn under the assumption of no contribution from astrophysical sources. 
While their purpose is mainly illustrative, they may not be unrealistic since
astrophysical sources, that are indeed required to account for the IGRB thanks to their medium-to-large redshift emission, can indeed provide a negligible contribution
to the cross-correlation signal between Fermi-LAT and 2MASS galaxies that, as we point out, has a rather
local origin (see discussion in Section S2 of the Supplemental Material~\cite{Supplemental}).
On the other hand, given the current uncertainty on the astrophysical components of the IGRB, an astrophysical model that can explain the measured cross correlation signal with no additional contribution from DM can be found~\cite{Xia:2015wka}.
Future data and analyses will help distinguishing between these two options.

This cross-correlation measurement can alternatively be used to derive 95\% C.L. upper bounds on the annihilation/decay rate.
These bounds are conservative and robust, since we assume here that DM is the only source of the \g-ray signal, without introducing additional assumptions on astrophysical components which  would make make the constraints stronger but also more model dependent. The 95\% C.L. upper bounds on the WIMP annihilation (decay) rate as a function of WIMP mass are shown in the left-hand (right-hand) panel of Fig.~\ref{fig:DMbounds}. For $b\bar b$ and $\tau^+\tau^-$ final states, the thermal annihilation rate is excluded for masses below 10 (100) GeV in the \low\ (\high) scenario. In the case of $\mu^+\mu^-$, the bounds degrade by about one order of magnitude.

\begin{figure}[t]
\vspace{-4cm}
\centering
\includegraphics[width=0.5\textwidth]{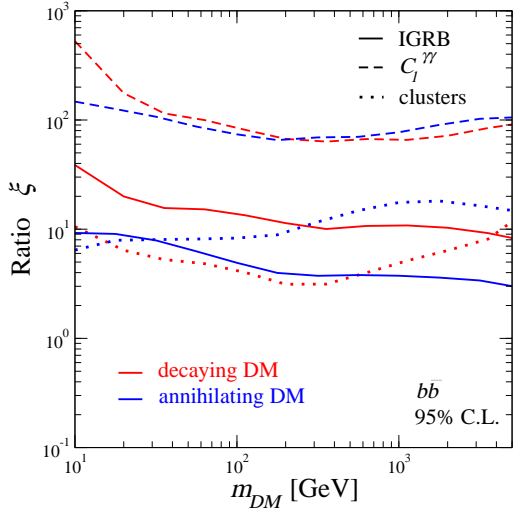}
\caption{Ratio $\xi=\sv_i/\sv_X$ (in the annihilating DM case) and $\xi=\tau_X/\tau_i$ (in the decaying DM case) between the 95\% C.L. bounds derived with method discussed in this work (cross-correlation of \g-rays with 2MASS catalogue, labeled with $X$) and from other extragalactic \g-ray probes ($i$ stands for total IGRB intensity, angular autocorrelation or clusters). The plot refers to the $b\bar b$ final state and (for the annihilating DM case) the \low\ substructure scheme.}
\label{fig:compar}
\end{figure}

In Fig.~\ref{fig:compar} we compare the sensitivity of our cross correlation method with that of other {\em extragalactic} \g-ray probes. 
We focus on these probes since they are similarly affected by uncertainties in modeling DM halo and sub halos properties. This allows to compare various techniques in a homogeneous and robust way, something that cannot be done with local DM tracers (Galactic regions, dwarf galaxies) or early Universe probes, which have different systematic uncertainties (see however the discussion in Section S1 of the Supplemental Material~\cite{Supplemental}). For illustrative purposes, we selected the \low\ substructure scheme and $b\bar b$ final states case. We verified that different choices provide little differences and the results are robust to both the DM clustering model and the annihilation/decay channel. 
We consider again the simplest case (where most conservative bounds can be derived), in which the astrophysical contribution is set to zero in all observables and only DM is contributing as \g-ray source.

The bound corresponding to the IGRB energy spectrum has been derived using the IGRB estimated by the Fermi-LAT Collaboration~\cite{Ackermann:2014usa} and adding up in quadrature statistical and systematic errors given in their Table 3. For the autocorrelation bound, we considered the angular spectrum estimated in four energy bins in Ref. \cite{Ackermann:2012uf} as provided in their Table II (DATA:CLANED) and averaged in the multipole range $155\leq\ell\leq 504$. For both probes, the model prediction has been computed using the same DM modeling as in our analysis.
Our bounds are compatible with the ones presented in Refs.~\cite{Ackermann:2015tah,Ajello:2015mfa,DiMauro:2015tfa,Ando:2015qda,Gomez-Vargas:2013cna,Ando:2013ff} (under the same set of assumptions). 
Cluster bounds are instead taken directly from the literature. In particular, for annihilating DM, we consider the analysis of 34 clusters using expected sensitivity for the 5 years Fermi-LAT data in Ref.~\cite{Zimmer:2015oka} which uses the same \low\ model adopted here. For decaying DM, we consider the analysis of 8 clusters in 3 years of Fermi-LAT data taking performed by~\cite{Huang:2011xr}.

Fig.~\ref{fig:compar} shows that the cross-correlation technique stands out as the most sensitive one, improving the constraints by a factor between a few to a hundred over the other techniques. Note that the ratio decreases at high energy because our analysis focuses at low energies (up to $E>10$ GeV). Since the IGRB is measured up to 820 GeV there is room for further improvements.

\section{Conclusions}
\label{sec:concl}
We compared the predicted angular cross-correlation between the \g-ray emission induced by DM annihilation or decay and the distribution of 2MASS galaxies with the measured CCF between these objects and the Fermi-LAT \g-ray maps. 

The contribution of astrophysical sources to the IGRB is assumed to be subdominant at low redshift and not included in the model prediction, so to derive conservative bounds on DM. We found that in the \low\ \cite{Sanchez-Conde:2013yxa} and \high\ \cite{Gao:2011rf} scenarios the ``thermal'' annihilation cross-section is excluded at 95\% C.L. up to DM masses of 10 and 100 GeV, respectively, for final state of annihilation into $b\bar b$ and $\tau^+\tau^-$.

We demonstrated that the cross-correlation technique is significantly more sensitive to a DM signal than all other {\em extragalactic} \g-ray probes used so far. This was done by comparing the bounds of our cross-correlation analysis with the most recent results from IGRB, angular autocorrelation and clusters, finding an improvement of a factor ranging from a few up to 100 for both annihilating and decaying DM.

We showed that a WIMP DM contribution can fully explain the observed cross correlation. A canonical WIMP with a mass in the 10--100 GeV range, annihilation rate around the thermal value, and realistic model for DM halo and sub-halo properties reproduce both size and shape of the measured angular cross-correlation. This intriguing possibility deserves further investigation within a more comprehensive framework that include contributions from astrophysical sources and additional data.

Future investigation employing the Pass8 release from the Fermi-LAT and forthcoming surveys at low/intermediate redshift~\cite{Bilicki14} will therefore provide remarkable insights to the particle DM quest.

\bigskip
\acknowledgments
This work is supported by the PRIN 2012 research grant {\sl Theoretical Astroparticle Physics} number 2012CPPYP7 funded by MIUR, by the research grants {\sl TAsP (Theoretical Astroparticle Physics)} and {\sl Fermi} funded by the INFN, and by the  {\sl Strategic Research Grant: Origin and Detection of Galactic and Extragalactic Cosmic Rays} funded by Torino University and Compagnia di San Paolo. JX is supported by the National Youth Thousand Talents Program, the National Science Foundation of China under Grant No. 11422323, and the Strategic Priority Research Program, {\sl The Emergence of Cosmological Structures} of the Chinese Academy of Sciences, Grant No. XDB09000000. MV and EB are supported by PRIN MIUR and IS PD51 INDARK grants. MV is also supported by ERC-StG cosmoIGM, PRIN INAF.

\setcounter{figure}{0}
\setcounter{equation}{0}
\setcounter{section}{0}
\setcounter{table}{0}
\renewcommand{\thesection}{S\arabic{section}}  
\renewcommand{\thetable}{S\arabic{table}}  
\renewcommand{\thefigure}{S\arabic{figure}}
\renewcommand{\theequation}{S\arabic{equation}}

{\center{\Large{\textsc{Supplemental Material}}}}

\section{Comparison with other \g-ray bounds}
When discussing bounds on WIMP DM, one typically refers to the constraints on its microscopic properties, i.e., on the mass and annihilation (decay) rate, for a given set of branching ratios of annihilation (decay) into ordinary particles.
Especially in the case of indirect signals, such bounds strongly depend on the DM distribution on different astrophysical scales: this often suffers of sizable uncertainties, which in turn reflect into uncertainties on the DM bounds.
For this reason, when comparing the results of this work with other bounds from the literature, we restricted our discussion in the main text to extragalactic probes: in fact, the bounds arising from the \g-ray average intensity, from the \g-ray angular autocorrelation and from galaxy clusters, have an origin which is (at least partially) common to the cross-correlations signal we discuss here, and therefore their magnitudes are expected to scale similarly when assuming different DM models.
For this reason we have included them explicitly in the main text.

In this Section, we attempt instead to compare our results with bounds that originates from \g-rays produced within or nearby our own Galaxy, focussing on annihilating DM for definiteness.
Since this comparison is more model dependent and much less straightforward, we define, for each probe, a conservative (CONS) and optimistic (OPT) scenario of DM spatial distribution (paralleling the \low\ and \high\ scenarios of the main analysis). We stress that this discussion only aims at providing a generic (although informative) comparison, whilst a comprehensive assessment of the various uncertainties in each single channel is clearly beyond the scope of the present analysis.

\begin{figure}[t]
\vspace{-3.5cm}
\centering
\includegraphics[width=0.54\textwidth]{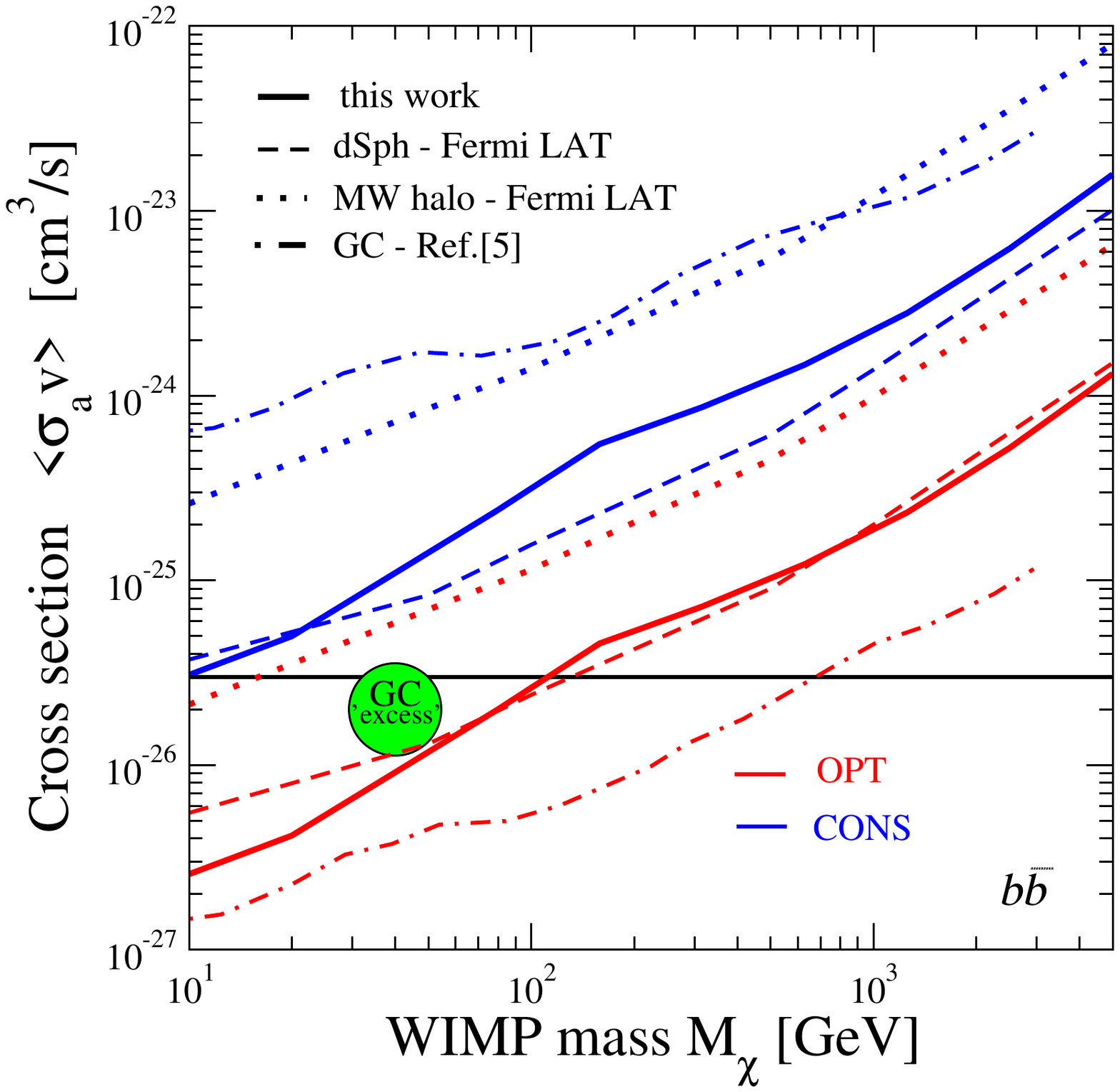}
\caption{Bounds on the DM annihilation rate as a function of its mass for various \g-ray probes, for a $b\bar b$ final state. The limits derived in this work from extragalactic DM signals are compared with the constraints from dSph galaxies~\cite{Ackermann:2015zua}, the MW halo~\cite{Ackermann:2012rg} and the GC~\cite{Gomez-Vargas:2013bea}. For all cases we define an optimistic (OPT) and conservative (CONS) scenario. The region compatible with the so-called GC excess~\cite{Daylan:2014rsa} is shown for comparison.}
\label{fig:DMboundsGAL}
\end{figure}

As mentioned, in the case of the angular cross-correlation between \g-rays from Fermi-LAT and 2MASS galaxies, the OPT and CONS scenarios are the \high\ and \low\ cases introduced in the main text. The associated 95\% C.L. bounds are reported with solid lines in Fig.~\ref{fig:DMboundsGAL} (taken from Fig.~3) for a $b\bar b$ final state.

Bounds from \g-ray emission in the halo of the Milky Way (MW) in Fig.~\ref{fig:DMboundsGAL} are taken from Ref.~\cite{Ackermann:2012rg}. To quantify uncertainties, we consider two different values of the local DM density, namely $\rho_0=0.7~ {\rm GeV\,cm^{-3}}$ (OPT) and $\rho_0=0.2~ {\rm GeV\,cm^{-3}}$ (CONS).
\footnote{The reported curves correspond to the $3\sigma$ bounds for an NFW profile, as shown in Fig.~4 of Ref.~\cite{Ackermann:2012rg}. Considering the 95\% C.L. limits or a different choice for the DM profile would not significantly alter the curves shown in Fig.~\ref{fig:DMboundsGAL}.}

Dwarf spheroidal (dSph) satellites of the MW are probably the objects that currently set most stringent \g-ray limits on DM.
The red dashed line in Fig.~\ref{fig:DMboundsGAL} is taken from the analysis of Ref. \cite{Ackermann:2015zua}, which
includes ultra-faint dSphs. Although the addition of the recently discovered ultra-faint dSphs can improve the DM bounds, the low number of detected stars in these systems makes the derivation of their dispersion velocity profile (and in turn of the DM content) quite uncertain.
Moreover, their proximity to the MW suggests strong tidal interactions which are normally not accounted for when solving the Jeans equation at the equilibrium to derive the DM profile.
A more conservative assessment of the uncertainties associated to the DM profile of ultra-faint dSph galaxies makes them less constraining than classical dSphs (see, e.g., Ref.~\cite{Bonnivard:2014kza,Bonnivard:2015gla}).
We thus include only the classical ones in the CONS scenario (and, conservatively, assume a Burkert profile which reduces the bounds by a factor of $\sim1.3$, see Fig.~7 of Ref.~\cite{Ackermann:2015zua}).
In order to re-derive the Fermi-LAT bounds for this scenario, we rescale the bound of Ref.~\cite{Ackermann:2015zua} by the ratio of the largest J-factors between the ultra-faint and classical dSphs. Although this rescaling is simplistic, nevertheless it grabs the essence of the uncertainty.
We also note here that other (typically ignored) uncertainties can affect the J-factor of the same classical dSphs by a factor of a few (e.g., systematics from triaxiality, see Ref.~\cite{Bonnivard:2014kza}): this implies that the blue dashed bound in Fig.~\ref{fig:DMboundsGAL} could be further pushed upward.
 
Finally, we consider the limits coming from the Galactic Center (GC).
They strongly depend on the assumption concerning the DM density in the innermost region of the MW, something which is not currently obtainable through observations.
This fact makes the bounds associated to the GC rather uncertain.
The dot-dashed curves shown in Fig.~\ref{fig:DMboundsGAL} are taken from Ref.~\cite{Gomez-Vargas:2013bea} which makes use of four years of Fermi-LAT data. The OPT scenario refers to a compressed NFW, while the CONS case to a canonical NFW. Note that they span nearly three orders of magnitude.
For comparison, we show also the preferred region for the so-called GC excess~\cite{Daylan:2014rsa}.

As for the DM bounds derived in the main text, and reproduced in Fig.~\ref{fig:DMboundsGAL}, we adopted the case where only DM is considered to contribute to the \g-rays signal.
In this way, the derived bounds are conservative, and no further assumption on the various possible astrophysical backgrounds (which would make the comparison more model dependent) is required. 
As clear from Fig.~\ref{fig:DMboundsGAL}, the bounds derived in this work are competitive with the most stringent Galactic bounds.
This also implies that a DM explanation of the cross-correlation signal (see Figs.~1 and 2 in the main text), or more generally DM scenarios lying in the region between the \high\ and \low\ constraints from cross-correlation, are not currently ruled out by other means.

We finally comment that we restricted the discussion to \g-ray bounds only.
We have not introduced other probes (such as charged cosmic-rays, radiative emissions, neutrinos, CMB) since they require further and completely different sets of assumptions (e.g., on the astrophysical environment and galactic transport) and one can always select a reasonable scenario such that the associated bounds become weaker than the ones shown in Fig.~\ref{fig:DMboundsGAL}
for the cross-correlation analysis.

In conclusion, the limits derived in the main text from the analysis of the cross-correlation 
signal between \g-rays extragalactic emission and large scale structure tracers are competitive and independent from 
the other bounds that can be derived from alternative means, and, as shown in the main text, represent the most constraining ones when compared to bounds arising from extragalactic probes.

\section{Discussion on $\gamma$-ray astrophysical backgrounds at low $z$}

The extragalactic \g-ray signal from annihilating or decaying DM peaks at very low redshift (in Fig.~S2, we show the contribution to the IGRB as a function of $z$ for the same annihilating DM \low\ case of Fig.~1; see, e.g., the discussion on DM window functions in Ref.~\cite{Fornengo:2013rga} for more details).
This implies that the size of the signal of its cross-correlation with a {\sl low redshift} catalogue is tightly connected to the total DM contribution to the IGRB.
This is not the case for more mundane astrophysical \g-rays sources, as already mentioned in the Introduction of the main text.
Indeed, the fraction of IGRB emitted by astrophysical sources at $z\lesssim0.1$ is below few percent of their total IGRB contribution, as can be clearly seen in Fig.~S2 (the models shown in Fig.~S2 are the same employed in Ref.~\cite{Fornengo:2014cya} and fit the whole Fermi-LAT IGRB measurement~\cite{Ackermann:2014usa}). 
This is a general property that holds for all the main extragalactic \g-ray source populations.

Therefore the tight link between cross-correlation signal and contribution to the IGRB that holds true for DM is not valid, at low redshifts, for
the astrophysical sources.
A large contribution to the IGRB might be accompanied by a low cross-correlation signal at low redshift, and a scenario where astrophysical sources make up the whole IGRB (shown in the inset of Fig.~1) with, on the other hand, a DM contribution dominating the cross-correlation with 2MASS (more in general, with a catalogue which peaks at very low redshifts, see e.g. Fig.~S2)   
is a viable situation. Two questions are at hand here: $i)$ whether such a scenario is reasonable from the point of view of properties of astrophysical \g-rays sources; $ii)$ whether this options is compatible with data. Let us start discussing the first point.

The vast majority of extragalactic sources present in the Fermi-LAT catalogue are blazars. This implies they are the \g-ray emitters whose average properties are best known.
As shown in Fig.~S2, the unresolved emission from BL Lacertae (BL Lac) and flat spectrum radio quasars (FSRQ) is strongly suppressed at low redshift.
Therefore blazars cannot explain the measured cross-correlation of Fermi-LAT data with 2MASS.

Two other extragalactic populations are thought to provide a large contribution to the IGRB: misaligned AGN (mAGN) and star forming galaxies (SFG).
Unfortunately, since a very low number of these objects have been detected so far by the Fermi-LAT telescope, the knowledge of their \g-ray luminosity function suffers of large uncertainties, even in the low redshift regime.
We cannot obtain firm predictions for the expected overall size of the associated cross-correlation signal.

Concerning the angular behaviour, let us note that, at scales $\lesssim 1^\circ$, the DM interpretation shown in Fig.~1 stems from the non-linear clustering.
At the redshift of 2MASS ($z\sim0.07$) such angular scales correspond to physical scales of $\sim$Mpc.
Therefore, a component providing a good fit to the data needs to have a power spectrum with a sizable non-linear term at Mpc scales, which disfavours an interpretation in terms of galactic objects.

\begin{figure}[t]
\vspace{-3.5cm}
\centering
\includegraphics[width=0.54\textwidth]{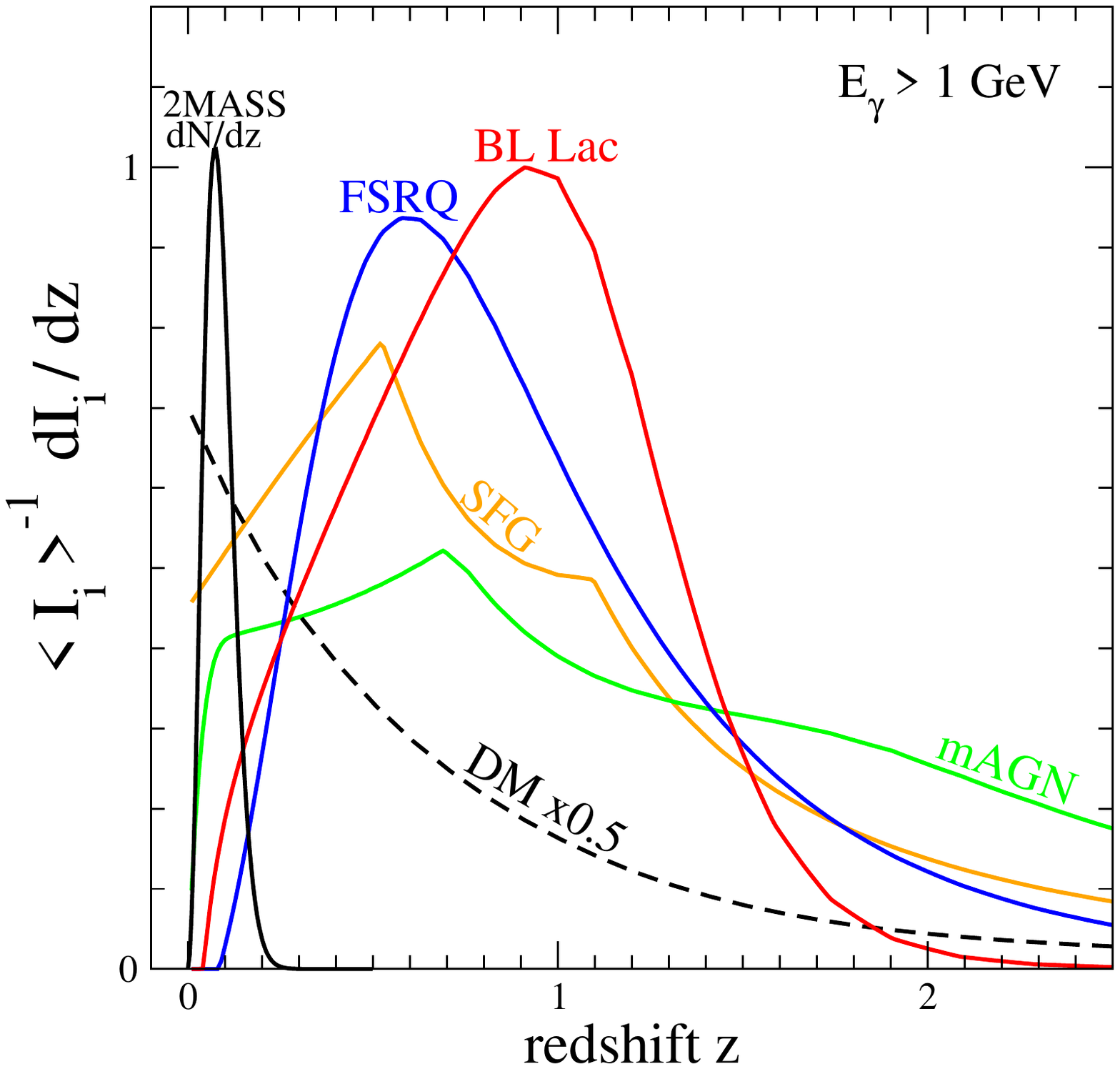}
\caption{Contribution to the IGRB for $E_\gamma>1$ GeV from various astrophysical \g-ray emitters as a function of redshift. The curves refer to the {\sl unresolved} component (we set the threshold for source detection to $F_{sens} = 5 \times 10^{-10}\, {\rm photons~ cm^{-2}~ s^{-1}}$). The black solid curve shows instead the redshift distribution of objects in the 2MASS catalogue.}
\label{fig:dIdz}
\end{figure}

In summary, there are currently significant uncertainties on the low-$z$ \g-ray luminosity function of the astrophysical components of the IGRB which prevent a robust determination of their contribution to the measured cross-correlation of Fermi-LAT data with 2MASS.

In the present analysis, we have been investigating the assumption of a subdominant astrophysical contribution combined with a particle DM explanation of the cross-correlation signal.
Confronting this scenario with data in the main section of the paper, the question $ii)$ have been addressed. We have demonstrated that it is, in principle, a viable possibility.

\section{Uncertainties from the choice of $M_{min}$ and $c(M)$}
For the sake of definiteness, in the main text, we fix the minimum halo mass to be $M_{min}=10^{-6} M_\odot$, and take the halo mass-concentration relation $c(M,z)$ from the reference case in Ref.~\cite{Sanchez-Conde:2013yxa} at masses above $10^{10} M_\odot$. The latter is then extrapolated to lower halo masses with two different prescriptions, \high\ and \low, as explained above.

In this Section, we explore the impact of a different choice for $M_{min}$ and of a different concentration model. 
In Fig.~S3, we show the clumping factor $\Delta^2$ introduced in Eq.~(2). It is a ``normalization'' of the DM signal which encodes the dependence on DM clustering properties.
The gray band shows the $\Delta^2$ region obtained by varying the $c(M)$ relation within the band provided in Ref.~\cite{Sanchez-Conde:2013yxa}.
Dotted lines are instead for $M_{min}=10^{-12} M_\odot$ (blue) and $M_{min}=10^{-2} M_\odot$ (red) (the fact they are close the borders of the gray band is just coincidental).
In this case, we focus on the \low\ clustering scenario, but similar scalings occur for the \high\ case as well.

Both effects introduce an extra factor of $\sim2$ of uncertainty in the predicted signal for annihilating DM. Such uncertainty band would translate into a corresponding uncertainty in $\sv$ of Fig.~\ref{fig:DMboundsGAL} (solid curves).

In the case of decaying DM, the normalization of the signal is instead not affected by the assumptions on $M_{min}$ and $c(M)$.

Above we just discussed the impact on the overall normalization of the signal.
Varying the $c(M)$ relation, the relative contributions of different halo masses would change and in turn the angular shape of the correlation can be affected.
This is however a minor effect (for reasonable $c(M)$) and is not relevant given the current accuracy of data.

\begin{figure}[t]
\vspace{-3.5cm}
\centering
\includegraphics[width=0.54\textwidth]{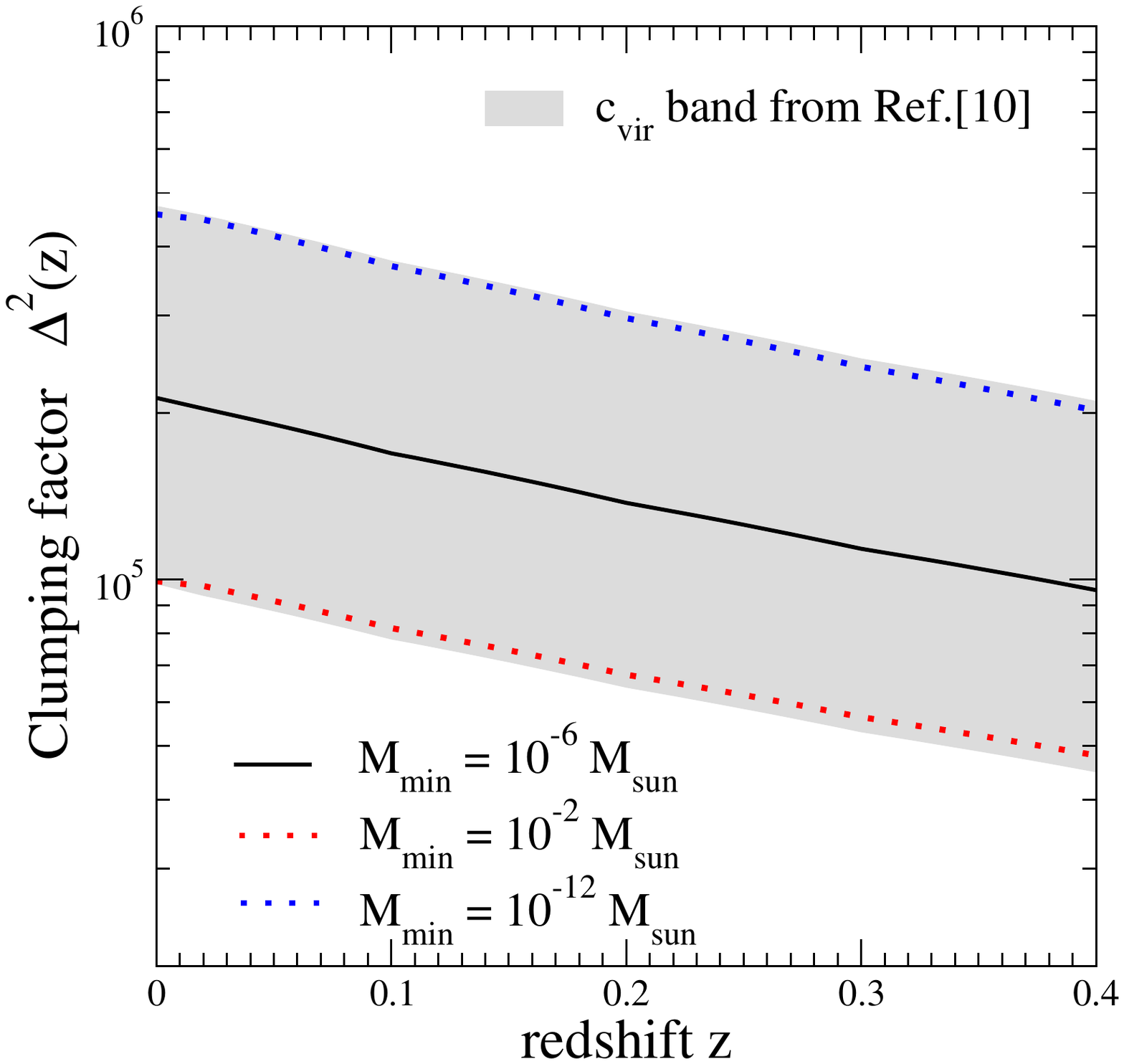}
\caption{Clumping factor $\Delta^2$ as a function of redshift for different choices of minimal halo mass $M_{min}$ and mass-concentration relation $c(M)$, as described in the labels.}
\label{fig:d2ave}
\end{figure}

\end{document}